\documentclass[mathleft
]{an}
\usepackage{graphicx}
\usepackage{times}
\newcommand{\teffq}{$T_{\!\mbox{\scriptsize \em eff}}^4$}
\newcommand{\teff}{$T_{\!\mbox{\tiny\rm eff}}$}

\newcommand{\zsun}{$Z_\odot$}
\newcommand{\msun}{$M_\odot$}
\newcommand{\ebv}{$E(B-V)$}

\newcommand{\hii}{H\,{\sc ii}\rm}

\newcommand{\oiii}{[O\,{\sc iii}]}
\newcommand{\oii}{[O\,{\sc ii}]}

\newcommand{\lin}{$\,\lambda$}

\begin{document}

\Pagespan{1}{}
\Yearpublication{2010}%
\Yearsubmission{2009}%
\Month{11}%
\Volume{999}%
\Issue{88}%

\title{Dissecting Galaxies with Quantitative Spectroscopy of the Brightest Stars in the Universe - Karl Schwarzschild Lecture 2009}

\author{Rolf-Peter Kudritzki\inst{1}\fnmsep\thanks{Corresponding author:
  \email{kud@ifa.hawaii.edu}\newline}
}
\titlerunning{Dissecting Galaxies}
\authorrunning{R. P. Kudritzki}
\institute{
Institute for Astronomy, University of Hawaii, 2680 Woodlawn Dr., Honolulu, Hawaii 96822, USA
}

\received{30 December 2009}
\accepted{10 January 2010}
\publonline{later}


\abstract{%
Measuring distances to galaxies, determining their chemical composition, 
investigating the nature of their stellar populations and the absorbing 
properties of their interstellar medium are fundamental activities in 
modern extragalactic astronomy helping to understand the evolution of 
galaxies and the expanding universe. The optically brightest stars in the 
universe, blue supergiants of spectral A and B, are unique tools for these 
purposes. With absolute visual magnitudes up to $M_{V} \cong -9.5$ they 
are the ideal to obtain accurate quantitative information about 
galaxies through the powerful modern methods of quantitative stellar 
spectroscopy. The spectral analyis of individual blue supergiant targets 
provides invaluable information about chemical abundances and abundance 
gradients, which is more comprehensive than the one obtained from \hii\ 
regions, as it includes additional atomic species, and which is also more accurate, 
since it avoids the systematic uncertainties inherent in the strong line 
studies usually applied to the \hii\ regions of spiral galaxies beyond the 
Local Group. Simultaneously, the spectral analysis yields stellar parameters 
and interstellar extinction for each individual supergiant target, which 
provides an alternative very accurate way to 
determine extragalactic distances through a newly developed method, 
called the Flux-weighted Gravity - Luminosity Relationship (FGLR). With 
the present generation of 10m-class telescopes these spectroscopic studies 
can reach out to distances of 10 Mpc. The new generation of 30m-class will 
allow to extend this work out to 30 Mpc, a substantial volume of the local 
universe.
}

\maketitle

\section{Introduction}

To measure distances to galaxies, to determine their chemical composition, and to investigate the nature of their 
stellar populations and the absorbing properties of their interstellar medium are fundamental 
activities in modern extragalactic astronomy. They are crucial to understand the evolution of galaxies 
and of the expanding universe and to constrain the history of cosmic chemical enrichment, from the metal-free 
universe to the present-day chemically diversified structure. However, while stars are the 
major constituents of galaxies, little of this activity is based on the quantitative 
spectroscopy of individual stars, a technique which over the last fifty years has proven to be one of the 
most accurate diagnostic tools in modern astrophysics. Given the distances to galaxies beyond 
the Local Group individual stars seem to be too faint for quantitative spectroscopy and, thus, 
astronomers have settled to restrict themselves to the photometric investigation of resolved 
stellar populations, the population synthesis spectroscopy of integrated stellar populations 
or the investigation of \hii\ region-emission lines. Of course, 
color-magnitude diagrams and the study of nebular emission lines have an impressive diagnostic 
power, but they are also limited in many ways and subject to substantial systematic uncertainties, 
as we will show later in the course of this lecture.

Thus, is it really out of the question to apply the methods of quantitative spectroscopy of 
individual stars as a most powerful complementary tool to understand the evolution of galaxies 
beyond the Local Group? The answer is, no, it is not. In the era of 10m-class telescopes with 
most efficient spectrographs it is indeed possible to quantitatively analyze the spectra of 
individual stars in galaxies as distant as 10 Mpc and to obtain invaluable information about 
chemical composition and composition gradients, interstellar extinction and extintion laws as 
well as accurate extragalactic distances. With the even larger and more powerful next generation 
of telescopes such as the TMT and the E-ELT we will be able to extend such studies out to distances 
as large as 30 Mpc. All one has to do is to choose the right type of stellar objects and to apply 
the extremely powerful tools of NLTE spectral diagnostics, which have already been successfully tested with high 
resolution, high signal-to-noise spectra of similar objects in the Milky Way and the Magellanic 
Clouds. 

Of course, now when studying objects beyond the Local Group the analysis methods need to be 
modified towards medium resolution spectra with somewhat reduced signal. For a stellar spectroscopist, 
who is usually trained to believe that only the highest resolution can give you an important answer,
this requires some courage and boldness (or maybe naive optimism). But as we will see 
in the course of this lecture, once one has done this step, a whole new universe is opening up in 
the true sense of the word.

\section{Choosing the Right Objects - A and B Supergiants}

It has long been the dream of stellar astronomers to study individual stellar 
objects in distant galaxies to obtain detailed spectroscopic information about 
the star formation history and chemodynamical evolution of galaxies and to
determine accurate distances based on the determination of stellar 
parameters and interstellar reddening and extinction. At the first glance, 
one might think that the most massive and, therefore, most luminous stars with 
masses higher than 50 $M_{\odot}$ are ideal for this purpose. However, because 
of their very strong stellar winds and mass-loss these objects keep very hot 
atmospheric temperatures throughout their life and, thus, waste most of their 
precious photons in the extreme ultraviolet. As we all know, most of these UV 
photons are killed by dust absorption in the star forming regions, where these 
stars are born, and the few which make it to the earth can only be observed with 
tiny UV telescopes in space such as the HST or FUSE and are not accessible to 
the giant telescopes on the ground.

Thus, one learns quickly that the most promising objects for such studies are massive 
stars in a mass range between 15 to 40 $M_{\odot}$ in the short-lived evolutionary
phase , when they leave the hydrogen main-sequence and cross the HRD in a few 
thousand to ten thousand years as blue supergiants of B and early A spectral type. Because of 
the strongly reduced absolute value of bolometric correction when evolving towards
smaller temperature these objects increase their brightness in visual light and 
become the optically brightest ``normal'' stars in the universe with absolute visual 
magnitudes up to $M_{V} \cong -9.5$ rivaling with the integrated light brightness 
of globular clusters and dwarf spheroidal galaxies. These are the ideal stellar 
objects to obtain accurate quantitative information about galaxies.

The optical spectra of B- and A-type supergiants are rich in metal absorption 
lines from several elements (C, N, O, Mg, Al, S, Si, Ti, Fe, among others). As 
young objects they represent probes of the current composition of the 
interstellar medium. Abundance determinations of these objects can therefore 
be used to trace the present abundance patterns in galaxies, with the ultimate 
goal of recovering their chemical and dynamical evolution history. In addition 
to the $\alpha$-elements, key for a comparison with \hii\ region results and to
understand the chemical evolutionary status of each individual star, the 
stellar analysis of blue supergiants provides the only accurate way to obtain 
information about the spatial distribution of Fe-group element abundances in 
external star-forming galaxies. So far, beyond the Local Group most of the 
information about the chemical properties of spiral galaxies has been 
obtained through the study of \hii\  regions oxygen emission lines using 
so-called strong-line methods, which, as we will show, have huge systematic uncertainties 
arising from their calibrations. Direct stellar abundance studies 
of blue supergiants open a completely new and more accurate way to investigate 
the chemical evolution of galaxies.

In addition, because of their enormous intrinsic brightness, blue supergiants are also ideal 
distance indicators. As first demonstrated by \cite{kud03} there is a very simple and compelling 
way to use them for distance determinations. Massive stars with masses in the range from 
12~\msun\ to 40~\msun\ evolve through the B and A 
supergiant stage at roughly constant luminosity. In addition, 
since the evolutionary timescale is very short when crossing through the B and A 
supergiant domain, the amount of mass lost in this stage is small. As a consequence,
the evolution proceeds at constant mass and constant luminosity. This has a very 
simple, but very important consequence for the relationship between gravity and 
effective temperature along each evolutionary track, namely that the flux-weighted gravity, 
$g_\mathrm{\scriptscriptstyle F}=g/T_\mathrm{eff}^4$, stays constant. As shown in detail by 
\cite{kud08} and explained again further below
this immediately leads to the \emph{`flux-weighted gravity--luminosity relationship' (FGLR)}, 
which has most recently proven to be an extremly powerful to determine extragalactic distances 
with an accuracy rivalling the Cepheid- and the TRGB-method. 

\section{Spectral Diagnostics and Studies in the Milky Way and Local Group}

The quantitative analysis of the spectra of these objects is not trivial. 
Most importantly, NLTE effects and but also the influence of stellar winds and atmospheric 
geometrical extension are of crucial importance. However, over the past decades with an effort of 
hundreds of men-years sophisticated model atmosphere codes for massive stars have been developed 
including the hydrodynamics of stellar winds and the accurate NLTE  opacities of millions of 
spectral lines. Detailed tests have been carried out reproducing the observed spectra of Milky Way 
stars from the UV to the IR and constraining stellar parameters with unprecedented accuracy (see 
reviews by \cite{kud98}, \cite{kud00}, \cite{kud09}). For instance, 
the most recent work on A-type supergiants (Przybilla et al.~2006, 2008, \cite{schiller08}, see 
also \cite{przybilla00}, \cite{przybilla01a}, \cite{przybilla01b}, \cite{przybilla01}, 
\cite{przybilla02}) demonstrates that with high resolution and very high signal-to-noise spectra 
stellar parameters and chemical abundances can be determined with hitherto unkown precision 
(\teff\/ to $\le 2$\%, $log~g$ to $\sim 0.05$ dex, individual metal abundances to $\sim 0.05$ dex).

At the same time, utelizing the power of the new 8m to 10m class telescopes, high resolution 
studies of A supergiants in many Local Group galaxies  were carried out by \cite{venn99} (SMC),
\cite{mccarthy95} (M33), \cite{mccarthy97} (M31), \cite{venn00} (M31),
\cite{venn01} (NGC 6822), \cite{venn03} (WLM), and \cite{kaufer04} (Sextans A) 
yielding invaluable information about the stellar chemical composition in these galaxies.
In the research field of massive stars, these studies have so far provided the most 
accurate and most comprehensive information about chemical composition and have been used 
to constrain stellar evolution and the chemical evolution of their host galaxies.

\section{The Challenging Step beyond the Local Group}

The concept to go beyond the Local Group and to study A supergiants by means of 
quantitative spectroscopy in galaxies out to the Virgo cluster has been first
presented by \cite{kud95} and \cite{kud98}. Following-up on this idea,
\cite{bresolin01} and \cite{bresolin02} used the VLT and FORS at 5 $\AA$ resolution
for a first investigation of blue supergiants in NGC 3621 (6.7 Mpc) and NGC 300 (1.9 Mpc). They were able
to demonstrate that for these distances and at this resolution spectra of sufficient S/N 
can be obtained allowing for the quantitative determination of stellar parameters 
and metallicities. \cite{kud03} extended this work and showed that stellar 
gravities and temperatures determined from the spectral analysis can be used to 
determine distances to galaxies by using the correlation between absolute bolometric
magnitude and flux weighted gravity $g_{F} = g/$\teffq\/ (FGLR).
However, while these were encouraging steps towards the use of A supergiants as quantitative
diagnostic tools of galaxies beyond the Local Group, the work presented in these papers
had still a fundamental deficiency. At the low resolution of 5 $\AA$ it is not possible
to use ionization equilibria for the determination of \teff\/ 
in the same way as in the high resolution 
work mentioned in the previous paragraph. Instead, spectral types were determined and
a simple spectral type - temperature relationship as obtained for the Milky Way was 
used to determine effective temperatures and then gravities and metallicities. Since
the spectral type - \teff\/ relationship must depend on metallicity (and also gravities),
the method becomes inconsistent as soon as the metallicity is significantly different from
solar (or the gravities are larger than for luminosity class Ia) and may lead to inaccurate
stellar parameters. As shown by  \cite{evans03}, the uncertainties introduced in this 
way could be significant and would make it impossible to use the FGLR for distance 
determinations. In addition, the metallicities derived might be unreliable. This posed
a serious problem for the the low resolution study of A supergiants in distant galaxies.

This problem was overcome only very recently by \cite{kud08} (herafter KUBGP), who provided the first 
self-consistent determination of stellar parameters and metallicities for A supergiants 
in galaxies beyond the Local Group based on the detailed quantitative model atmosphere 
analysis of low resolution spectra. They applied their new method on 24 supergiants of 
spectral type
B8 to A5 in the Scultor Group spiral galaxy NGC 300 (at 1.9 Mpc distance) and obtained 
temperatures, gravities, metallicities, radii, luminosities and masses. The spectroscopic 
observations were obtained with FORS1 at the ESO VLT in multiobject 
spectroscopy mode. In addition, ESO/MPI 2.2m WFI and HST/ACS photometry was used. The observations 
were carried out within the framework of the Araucaria Project (\cite{gieren05b}).
In the following we discuss the analysis method and the results of this pilot study.

\section{A Pilot Study in NGC300 - The Analysis Method}

For the quantitative analysis of the spectra KUBGP use the same combination
of line blanketed model atmospheres and very detailed NLTE line formation calculations
as \cite{przybilla06} in their high signal-to-noise and high spectral resolution
study of galactic A-supergiants, which reproduce the observed normalized spectra and 
the spectral energy distribution, including the Balmer jump, extremely well. 
They calculate an extensive, comprehensive and dense grid of model atmospheres and NLTE 
line formation covering the potential full parameter range of all the objects in 
gravity ($log~g$ = 0.8 to 2.5), effective temperature (\teff = 8300 to 15000K) and 
metallicity ([Z] = log{Z/\zsun} = -1.30 to 0.3). The total grid comprises more than 6000 
models.

\begin{figure}[t]
\begin{center}
 \includegraphics[width=\linewidth]{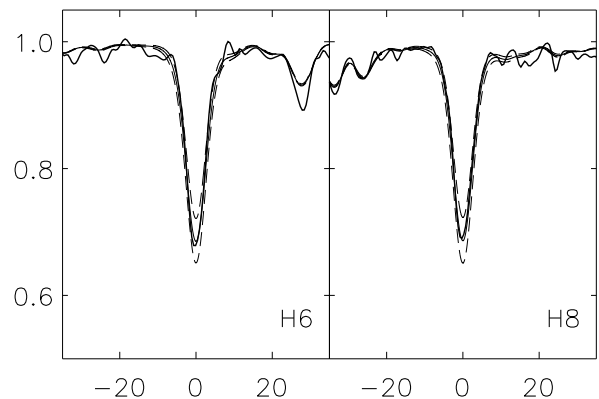} 
\caption{Model atmosphere fit of two observed Balmer lines of 
NGC300 target No. 21 of KUBGP for \teff\/ = 10000~K and $log~g$ = 1.55 (solid). 
Two additional models with same \teff\/ but $log~g$ = 1.45 and 1.65, 
respectively, are also shown (dashed).}
   \label{fig1}
\end{center}
\end{figure}

\begin{figure}[t]
\begin{center}
 \includegraphics[width=0.73\linewidth, angle=90]{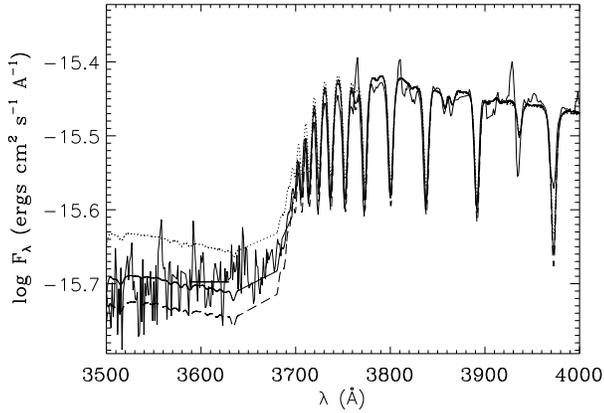} 
\caption{Model atmosphere fit of the observed Balmer jump of the same 
target as in Fig.\,\ref{fig1} for
\teff\/ = 10000~K and $log~g$ = 1.55 (solid). Two additional models with the same $log~g$
but \teff\/ = 9750~K (dashed) and 10500~K (dotted) are also shown. The horizontal bar at 
3600~\AA~represents the average of the flux logarithm over this wavelength interval, which 
is used to measure D${_B}$.}
   \label{fig2}
\end{center}
\end{figure}

The analysis of the each of the 24 targets in NGC 300 proceeds in three steps. First, 
the stellar 
parameters (\teff\/ and $log~g$) are determined together with interstellar reddening 
and extinction, then the metallicity is determined  and finally, assuming a distance to 
NGC 300, stellar radii, luminosities and masses are obtained. For the first step,
a well established method to obtain the stellar parameters of supergiants of late B to 
early A spectral type is to use ionization equilibria of weak metal lines (OI/II; MgI/II;
NI/II etc.) for the determination of effective temperature \teff\/ and the Balmer lines 
for the gravities $log~g$. However, at the low resolution
of 5 \AA~the weak spectral lines of the neutral species disappear in the noise of the 
spectra and an alternative technique is required to obtain temperature information. 
KUBGP confirm the result by \cite{evans03} that a simple application of a spectral 
type - effective temperature relationship does not work because of the degeneracy of 
such a relationship with metallicity. Fortunately, a way out of this dilemma is the 
use of the spectral energy distributions (SEDs) and here, in particular of the Balmer 
jump $D_B$. While the observed photometry from B-band to I-band is 
used to constrain the interstellar reddening, $D_B$ turns out to be a reliable 
temperature diagnostic. A simultaneous fit of the 
Balmer lines and the Balmer jump allows to constrain effective temperature and gravity 
independent of assumptions on metallicity. Figure\,\ref{fig1} and Fig.\,\ref{fig2} 
demonstrate the sensitivity of the Balmer lines and the Balmer jump to gravity and 
effective temperature, respectively. The accuray obtained by this method is $\le 4$\% 
for \teff\ and $\sim 0.05$ dex for $log~g$.

\begin{figure}[t]
\vskip-3mm
\hspace*{-0.7 cm}
 \includegraphics[width=1.0\linewidth]{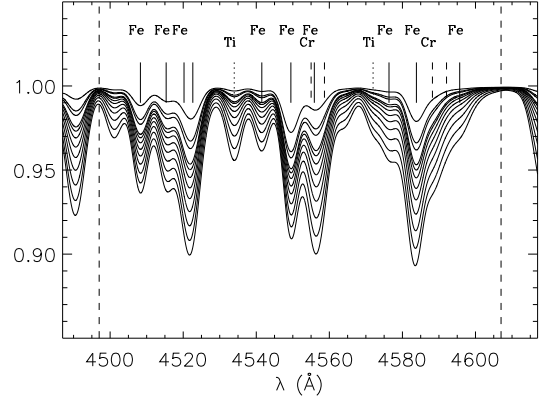} 
 \caption{Synthetic metal line spectra calculated for the stellar parameters of target 
No.21 as a function of metallicity in the spectral window from 4497~\AA~to 4607~\AA. 
Metallicities range from [Z] = -1.30 to 0.30, as described in the text. The dashed 
vertical lines give the edges of the spectral window as used for a determination of 
metallicity.}
   \label{fig3}
\end{figure}

Knowing the stellar atmospheric parameters \teff\/ and $log~g$ KUBGP are able to 
determine stellar metallicities by fitting the metal lines with their 
comprehensive grid of line formation calculations. The fit procedure proceeds again
in several steps. First, spectral windows are defined, for which a good definition 
of the continuum is possible and which are relatively undisturbed by flaws in 
the spectrum (for instance caused by cosmic events) or interstellar emission 
and absorption. A typical spectral window used for all targets is the 
wavelength interval 4497~\AA\/  $\le \lambda \le$ 4607~\AA. Figure\,\ref{fig3} shows 
the synthetic spectrum calculated for the atmopsheric parameters of the same target as analyzed in Fig. 1 and 2 
for all the metallicities 
of the grid ranging from -1.30 $\le$ [Z] $\le$ 0.30. It is very obvious that the 
strengths of the metal line features are a strong function of metallicity.

\begin{figure*}[t]
 \hspace*{-1.0 cm}
 \includegraphics[width=\textwidth]{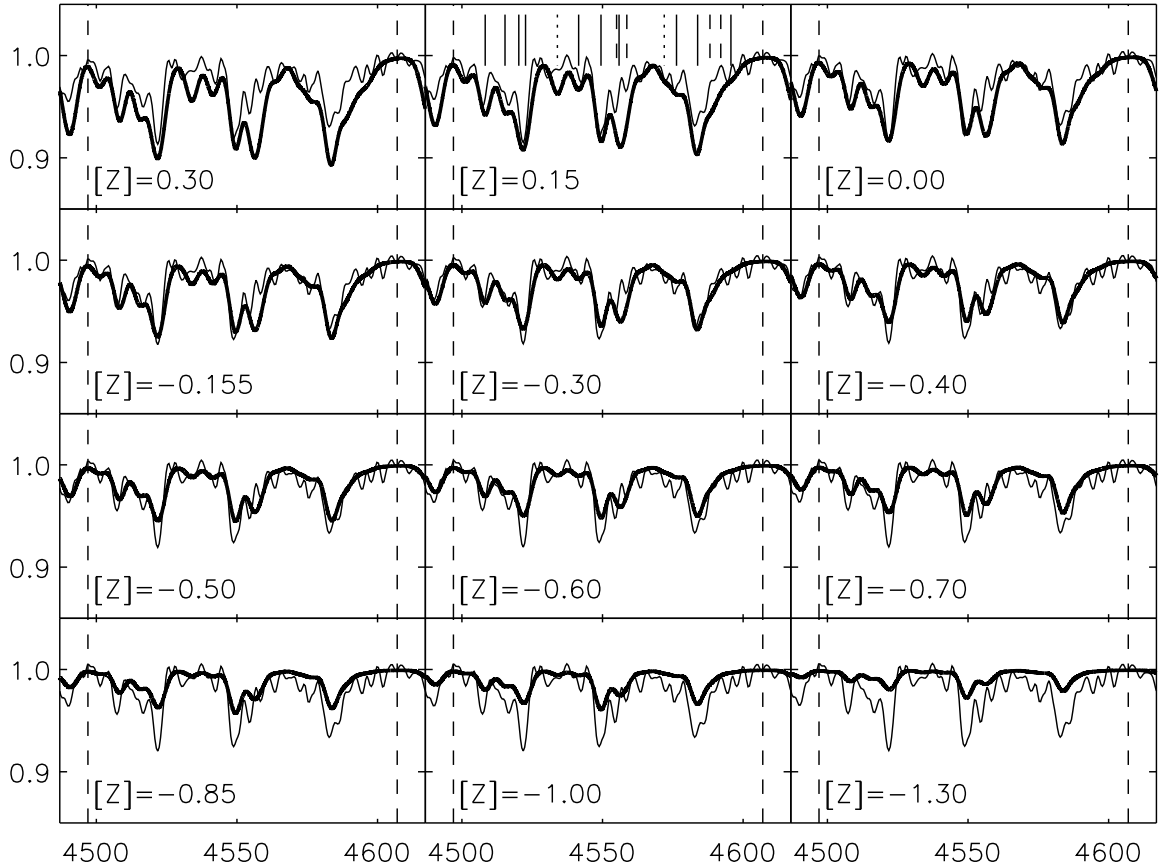} 
\caption{Observed spectrum of the same target as in Fig. 1 and 2 in the same spectral window as 
Fig.\,\ref{fig3} but now the synthetic spectra for each metallicity overplotted 
separately.}
   \label{fig4}
\end{figure*}

\begin{figure}[t]
\hskip-5mm
 \includegraphics[width=\linewidth]{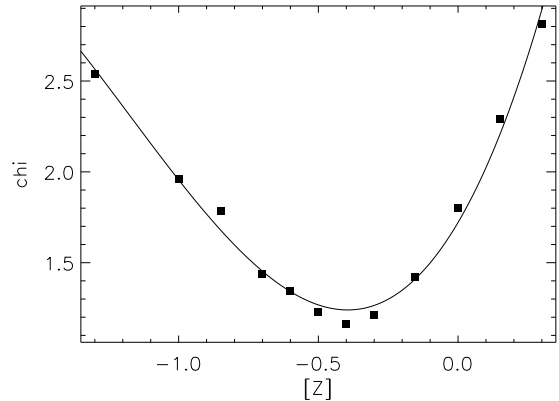} 
\caption{$\chi ([Z])$ as obtained from the comparison of observed and 
calculated spectra. The solid curve is a third order polynomial fit.}
   \label{fig5}
\end{figure}

In 
Fig.\,\ref{fig4} the observed spectrum of the target in this spectral 
window is shown overplotted by the synthetic spectrum for each metallicity. Separate 
plots are used for each metallicity, because the optimal relative normalization of the observed and 
calculated spectra is obviously metallicity dependent. This problem is addressed by 
renormalizing the observed spectrum for each metallicity so that the synthetic 
spectrum always intersects the observations at the same value at the two edges of 
the spectral window (indicated by the dashed vertical lines).  
The next step is a pixel-by-pixel comparison of calculated and normalized observed 
fluxes for each metallicity and a calculation of a $\chi^{2}$-value. The minimum 
$\chi ([Z])$ as a function of [Z] is then used to determine the metallicity. This 
is shown in Fig.\,\ref{fig5}. Application of the same method on different spectral
windows provides additional independent information on 
metallicity and allows to determine the average metallicity 
obtained from all windows. A value of [Z] = -0.39 is found with a very small dispersion of only 
0.02 dex. However, one also need to consider the 
effects of the stellar parameter uncertainties on the metallicity determination. This 
is done by applying the same correlation method for [Z] for models at the extremes 
of the error box for \teff\/ and $log~g$. This increases the uncertainty of [Z] 
to $\pm$ 0.15 dex, still a very reasonable accuracy of the abundance determination.

After the determination of \teff, $log~g$, and [Z], the model atmosphere SED is 
used to determine interstellar reddening E(B-V) and extinction A$_{V}$ = 3.1 E(B-V). 
Simultaneously, the fit also yields the stellar angular diameter, which provides 
the stellar radius, if a distance is adopted. 
\cite{gieren05} in their multi-wavelength study of a large sample 
of Cepheids in NGC 300 including the near-IR have determined a new distance 
modulus m-M = 26.37 mag, which corresponds to a distance of 1.88 Mpc. KUBGP have 
adopted these values to obtain the radii and absolute magnitudes. 

\begin{figure}[t]
\begin{center}
 \includegraphics[width=0.70\linewidth, angle=90]{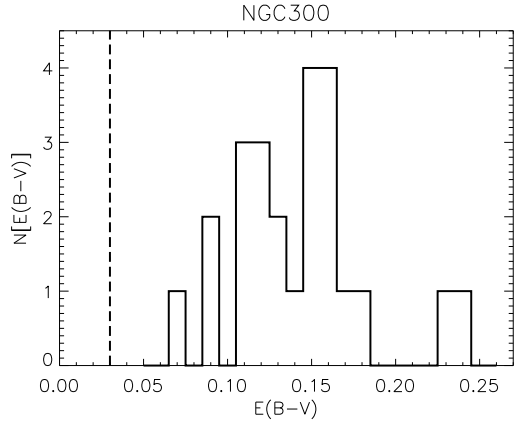} 
\caption{Histogram of the reddening distribution of blue supergiants in NGC~300 
determined from spectral analysis and photometry. The dashed lines show the \ebv\ 
value adopted by the {\em HST} Key Project. Data from Kudritzki et al.~(2008)}
   \label{fig6}
\end{center}
\end{figure}

\section{ A Pilot Study in NGC300 - Results}

As a first result, the quantitative spectroscopic method yields interstellar 
reddening and extinction as a by-product of the analysis process. For objects 
embedded in the dusty disk of a star forming spiral galaxy one expects a wide 
range of interstellar reddening E(B-V) and, indeed, a range from E(B-V) = 
0.07 mag up to 0.24 mag was found (see Fig.\,\ref{fig6}). The individual reddening values are 
significantly larger than the value of 0.03 mag 
adopted in the HST distance scale key project study of Cepheids by \cite{freedman01}
and demonstrate the need for a reliable reddening
determination for stellar distance indicators, at least as long the study is 
restricted to optical wavelengths. The average over the observed sample is 
$\langle E(B-V) \rangle $ = 0.12 mag in close agreement with the value of 0.1 mag 
found by \cite{gieren05} in their optical to near-IR study of Cepheids in NGC 300. 
While Cepheids have somewhat lower masses than the A supergiants of our study and 
are consequently somewhat older, they nonetheless belong to the same population and 
are found at similar sites. Thus, one expects them to be affected by interstellar 
reddening in the same way as A supergiants. 

\begin{figure}[t]
\begin{center}
 \includegraphics[width=0.70\linewidth, angle=90]{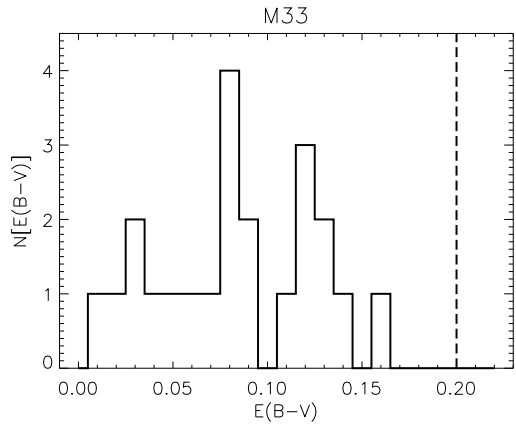} 
\caption{Histograms of the reddening  distribution of blue supergiants in M33. 
The dashed lines show the \ebv\ value adopted by the {\em HST} Key Project. Data from 
U et al.~(2009).}
   \label{fig7}
\end{center}
\end{figure}

\begin{figure}[t]
\begin{center}
 \includegraphics[width=0.70\linewidth, angle=90]{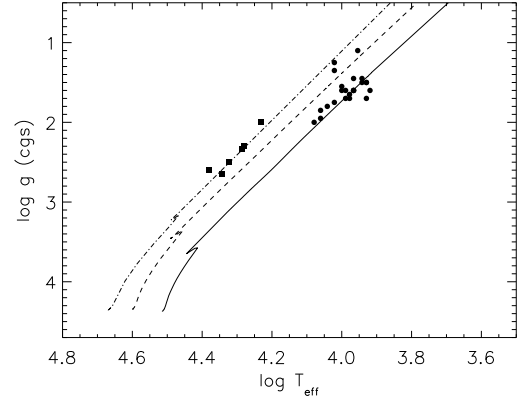} 
\caption{NGC 300 A supergiants (filled circles) and early B supergiants (filled squares)
 in the $(log~g, log$~\teff) plane compared with evolutionary 
 tracks by \cite{meynet05} of stars with 15 M$_{\odot}$ (solid), 25 M$_{\odot}$ 
(dashed), and 40 M$_{\odot}$ (dashed-dotted), respectively.
}
   \label{fig8}
\end{center}
\end{figure}

Note that a difference of 0.1 mag in 
reddening corresponds to 0.3 mag in the distance modulus. It is, thus, not surprising 
that \cite{gieren05} found a signicantly shorter distance modulus for NGC 300 than the 
Key Project. Another illustrative example of the importance of an accurate 
determination of interstellar extinction
is given in Fig.\,\ref{fig7}, which compares the reddening distribution obtained from 
the quantitative spectral analysis of blue supergiants in M33 with the value adopted 
by the Key Project. Again a wide distribution in reddenening is found, but this time 
the values are significantly smaller than the Key Project value (see discussion in 
section 9). 

\begin{figure}[t]
\begin{center}
 \includegraphics[width=0.70\linewidth, angle=90]{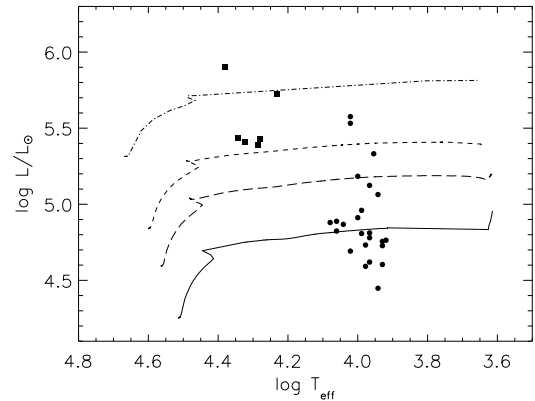} 
\caption{NGC 300 A and early B supergiants in the HRD compared with evolutionary 
 tracks for stars with 15 M$_{\odot}$ (solid), 
20 M$_{\odot}$ (long-dashed), 25 M$_{\odot}$ (short-dashed), 
and 40 M$_{\odot}$ (dashed-dotted), respectively.
}
   \label{fig9}
\end{center}
\end{figure}

Figure\,\ref{fig8} and Fig.\,\ref{fig9}  shows the location of all the observed targets in the 
$(log~g, log$~\teff) plane and in the HRD. The diagrams include the early B-supergiants studied by
\cite{urbaneja05b}. The comparison with evolutionary tracks gives a first 
indication of the stellar masses in a range from 10 M$_{\odot}$ to 40 M$_{\odot}$.
Three A supergiant targets have obviously higher masses than the rest of the sample and seem to be
on a similar evolutionary track as the objects studied by \cite{urbaneja05b}.
The evolutionary information obtained from the two diagrams appears to be 
consistent. The B-supergiants seem to be more massive than most of the A 
supergiants. The same three A supergiants apparently more massive than the rest 
because of their lower gravities are also the most luminous objects. 
This confirms that quantitative spectroscopy is -at least qualitatively - capable 
to retrieve the information about absolute luminosities. Note that the fact that 
all the B supergiants studied by \cite{urbaneja05b} are more massive is simply 
a selection effect of the V magnitude limited spectroscopic survey by 
\cite{bresolin02}. At similar V magnitude as the A supergiants those objects 
have higher bolometric corrections because of their higher effective temperatures 
and are, therefore, more luminous and massive.  

\begin{figure}[t]
\begin{center}
 \includegraphics[width=0.70\linewidth,angle=90]{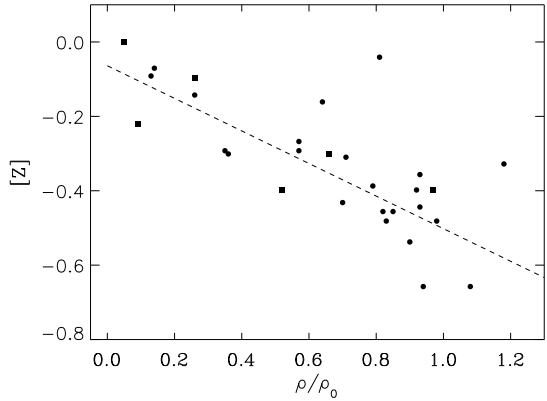} 
 \caption{Metallicity [Z] as a function of angular galacto-centric distance 
$\rho/\rho_{0}$ for the A supergiants (filled circles) and
early B-supergiants (filled squares). Note 
that for the latter metallicity refers to oxygen only. The dashed curve 
represents the regression discussed in the text.
}
   \label{fig10}
\end{center}
\end{figure}

Figure\,\ref{fig10} shows the stellar metallicities and the metallicity gradient as
a function of angular galactocentric distance, expressed in terms of the isophotal radius, 
$\rho/\rho_{0}$ ($\rho_{0}$ corresponds to 5.33 kpc). Despite the scatter 
caused by the metallicity uncertainties of the 
individual stars the metallicity gradient of the young disk population in NGC 300 
is very clearly visible. A linear regression for the combined A- and B-supergiant 
sample yields (d in kpc, see also \cite{bresolin09b})
\begin{equation}
        [Z] = -0.07\pm{0.05} - (0.081\pm{0.011})~d.
   \end{equation}

Note that the metallicities of the B supergiants 
refer to oxygen only with a value of log N(O)/N(H) = -3.35 adopted for the sun 
(\cite{asplund05}). On the other hand, the A supergiant metallicities reflect 
the abundances of a variety of heavy elements such as Ti, Fe, Cr, Si, S, and Mg.
KUBGP discuss the few outliers in Fig.\,\ref{fig10} and claim that these metallicities 
seem to be real. Their argument is that the expectation of homogeneous azimuthal 
metallicity in patchy star forming galaxies seems to be naive. Future work on other 
galaxies will show whether cases like this are common or not.

\begin{figure}[t]
\begin{center}
 \includegraphics[width=\linewidth]{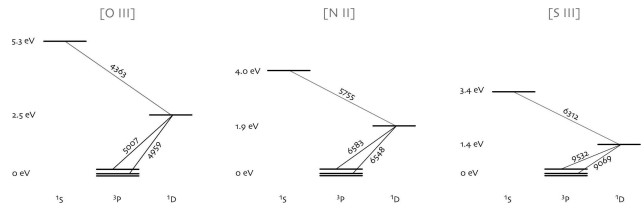} 
 \caption{Energy level diagrams for the strong and auroral nebular forbidden 
          lines. The auroral lines arise from the higher excited levels.
          (Bresolin, priv. comm.)}
   \label{fig11}
\end{center}
\end{figure}

It is important to compare these results with metallicity measurements 
obtained from \hii\ region emission lines. Beyond the Local Group and for metal 
rich galaxies \hii\ region metallicities are mostly restricted to oxygen and measured 
through the so-called 'strong-line method', which uses the fluxes of the strongest 
forbidden lines of \oii\ and \oiii\ relative to H$\beta$. Unfortunately, abundances 
obtained with the strong-line method depend heavily on the calibration used. Consequently,
for the comparison of stellar with \hii\ region metallicities KUBGP (extending the 
discussion started by \cite{urbaneja05b}) used line fluxes published by \cite{deharveng88} 
and applied various different published strong line method calibrations to determine nebular
oxygen abundances, which could then be used to obtain the similar regressions as 
above. 

\begin{figure}[t]
\begin{center}
 \includegraphics[width=0.97\linewidth]{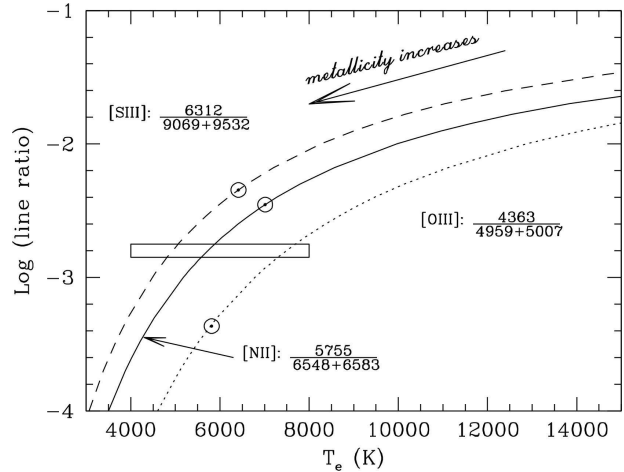} 
 \caption{Diagnostic of \hii\ regions. Ratio of auroral to strong line fluxes 
           as a function of nebular electron temperature. The horizontal box 
           indicates the observational limit for 10m-class telescopes. 
           (Bresolin, priv. comm.)}
   \label{fig12}
\end{center}
\end{figure}

As a very disturbing result, the different strong line method calibrations lead to 
significant differences in 
the central metallicity as well as in the abundance gradient. The calibrations 
by \cite{dopita86} and \cite{zaritsky94} predict a metallicity significantly 
supersolar in the center of NGC 300 contrary to the other calibrations. On the other hand, the work by KUBGP 
yields a central metallicity slightly smaller than solar in good agreement with 
\cite{denicolo02} and marginally agreeing with \cite{kobulnicky99}, 
\cite{pilyugin01}, and \cite{pettini04}. At the isophotal radius, 5.3 kpc 
away from the center of NGC 300, KUBGP obtain an average metallicity significantly 
smaller than solar [Z] = -0.50, close to the average metallicity in the SMC. The 
calibrations by \cite{dopita86}, \cite{zaritsky94}, \cite{kobulnicky99} 
do not reach these small values for oxygen in the HII regions either because their 
central metallicity values are too high or the metallicity are gradients too shallow.

\begin{figure}[t]
\begin{center}
 \includegraphics[width=0.98\linewidth]{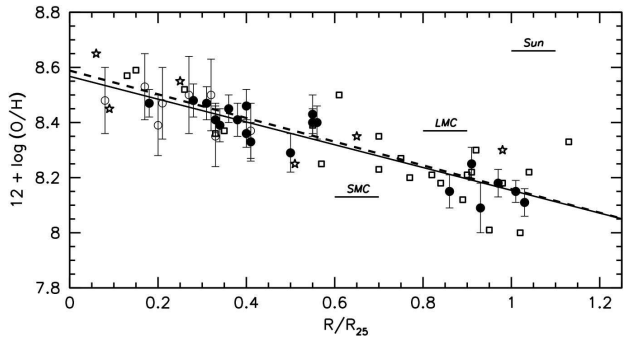} 
 \caption{Radial oxygen abundance  gradient obtained from \hii\ regions (circles) 
and blue supergiants (star symbols: B supergiants; open squares: A supergiants). The 
regression to the \hii\ region data is shown by the continuous line. The dashed line 
represents the regression to the BA supergiant star data. For reference, the oxygen 
abundances of the Magellanic Clouds (LMC, SMC) and the solar photosphere are 
marked. From \cite{bresolin09a}.}
   \label{fig13}
\end{center}
\end{figure}

\begin{table*}[ht]
\caption{Central metallicity [Z] and metallicity gradient (dex/kpc) 
         in NGC 300.}
\begin{tabular}{l c c c}
\hline\noalign{\smallskip}
Source & Central Abundance & Gradient & Comments \\ [0.5ex]
\hline
\\[-1mm]
Dopita \& Evans(1986)     &  0.29$\pm{0.17}$ & -0.118$\pm{0.019}$ & HII, oxygen  \\
Zaritsky et al. (1994)    &  0.32$\pm{0.04}$ & -0.101$\pm{0.017}$ & HII, oxygen  \\
Kobulnicky et al. (1999)  &  0.09$\pm{0.04}$ & -0.051$\pm{0.017}$ & HII, oxygen  \\
Denicolo et al. (2002)    & -0.05$\pm{0.05}$ & -0.086$\pm{0.019}$ & HII, oxygen  \\
Pilyugin (2001)           & -0.14$\pm{0.06}$ & -0.053$\pm{0.023}$ & HII, oxygen  \\
Pettini \& Pagel (2004)   & -0.16$\pm{0.04}$ & -0.068$\pm{0.015}$ & HII, oxygen  \\
KUBGP                     & -0.07$\pm{0.09}$ & -0.081$\pm{0.011}$ & stars, metals\\
\hline
\end{tabular}
\label{tab1}
\end{table*}

Thus, which of the metallicities determined for NGC 300 are correct? Those obtained 
from the blue supergiant study or the ones obtained from HII regions with one of the 
strong line calibrations? One possible way out of this problem is the use of the faint 
auroral emission lines (e.g.~\oiii\lin4363) of \hii\  regions, instead of relying 
solely on strong lines, for nebular metallicity determinations (see Fig.\,\ref{fig11}). This requires 
a substantially larger observational effort at 10m-class telescopes. \cite{bresolin09a} have recently used FORS at the VLT and studied 
28 \hii\ regions in NGC~300, for all of which they were able to detect the auroral 
lines and to use them to constrain nebular electron temperatures (see Fig.\,\ref{fig12}). This allowed for a 
much more accurate determination of nebular oxygen abundances, avoiding the calibration 
uncertainty intrinsic to the strong-line method. The result of this work compared to the 
blue supergiant metallicities is shown in Fig.\,\ref{fig13}. The agreement between the 
stars and the \hii\ regions is excellent, confirming independently the quality of the 
supergiant work and ruling out many of the strong line calibrations.

\begin{figure}[t]
\begin{center}
 \includegraphics[width=0.98\linewidth]{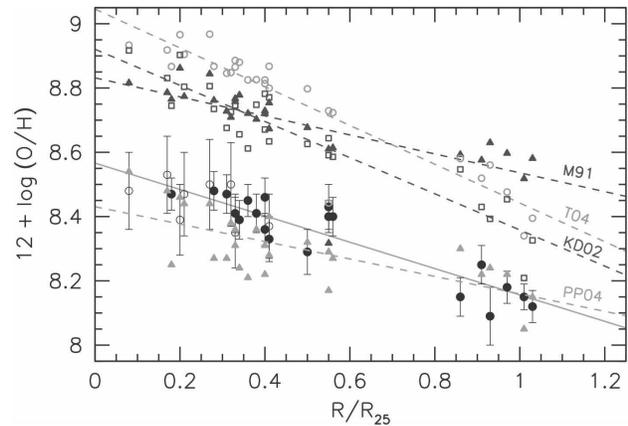} 
\caption{\hii\ region galactocentric oxygen abundance gradients in NGC~300 obtained from 
our dataset but different strong line  calibrations: 
 McGaugh~1991\,=\,M91, Tremonti et al.~2004\,=\,T04, Kewley \& Dopita~2002\,=\,KD02, 
and Pettini \& Pagel~2004\,=\,PP04, as shown by the labels to the corresponding 
least squares fits. The auroral line-based abundances determined 
by Bresolin et al.~(2009a) are shown by the full and open circle symbols, and 
the corresponding linear fit is shown by the continuous line.
}
   \label{fig14}
\end{center}
\end{figure}

\cite{bresolin09a} also perform a very educative experiment. They use their new and very 
accurate measuremants of strong line fluxes and simply apply a set of different more 
recent strong line calibrations to obtain oxygen abundances without using the 
information from the auroral lines. The result of this experiment compared with the 
oxygen abundances using the auroral lines is shown in Fig.\,\ref{fig14}. The comparison 
is again shocking. The abundance offsets introduced by the application 
of inappropriate strong-line calibrations can be as large as 0.6 dex, putting 
the whole business of constraining galaxy evolution through the measurement of nebular 
metallicities and metallicity gradients into jeopardy. 

\section{The Metallicities of Galaxies}

\begin{figure}[t]
\begin{center}
 \includegraphics[width=0.97\linewidth]{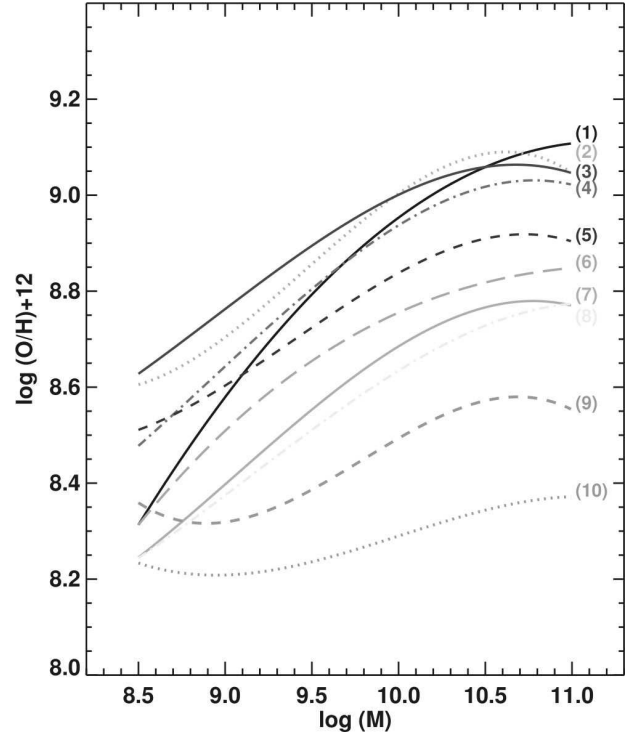} 
\caption{\small The mass-metallicity relationship of star forming galaxies in the nearby 
universe obtained by applying several widely used empirical metallicity calibrations based on 
different strong line ratios. This figure illustrates that there is an effect not only on the 
absolute scale, but also on the relative shape of this relationship. 
Adapted from Kewley \& Ellison~(2008).}
   \label{fig15}
\end{center}
\end{figure}

As is well known, the metallicity of the young stellar population of spiral galaxies 
has the potential to provide 
important constraints on galactic evolution and the chemical evolution history
of the universe. For instance, the relationship between central metallicity and galactic 
mass appears to be a Rosetta stone to understand chemical evolution and 
galaxy formation (\cite{lequeux79}, \cite{tremonti04}, \cite{maiolino08}). 
In addition, the observed metallicity gradients in spiral galaxies, 
apparently large for spirals of lower mass and shallow for high mass galaxies 
(\cite{garnett97}, \cite{skillman98}, \cite{garnett04}), are the result of a complex interplay of star formation 
history, initial mass function and matter infall into the disks of spirals and allow in 
principle to trace the evolutionary history of spiral galaxies.
However, as intriguing the observations of the mass-metallicity relationship 
and the metallicity gradients of galaxies are, the published results are 
highly uncertain in a quantitative sense, since they reflect the intrinsic uncertainty 
of the calibration of the strong line method applied. As a striking example, \cite{kewley08} 
have demonstrated that the quantitative shape of the mass-metallicity 
relationship of galaxies can change from very steep to almost flat depending 
on the calibration used (Fig.~\ref{fig15}). In the same way and as demonstrated by Fig.\,\ref{fig14}, and Tab.\,\ref{tab1} 
metallicity gradients of spiral galaxies can change from steep to flat simply as the result of the calibration used.
Obviously, the much larger effort of either stellar spectroscopy of blue supergiants or the 
observation of faint nebular auroral lines will be needed in the future to observationally 
constrain the metallicities and metallicty gradients of spiral galaxies and their evolutionary 
history. We note that the use of blue supergiants provides the additional information about 
chemical abundances of iron-group elements as an important additional constraint of galaxy 
evolution history.

\section{Flux Weighted Gravity - Luminosity Relationship (FGLR)}

Massive stars with masses in the range from 
12 M$_{\odot}$ to 40 M$_{\odot}$ evolve through the B and A 
supergiant stage at roughly constant luminosity (see Fig.\,\ref{fig5}). In addition, 
since the evolutionary timescale is very short when crossing through the B and A 
supergiant domain, the amount of mass lost in this stage is small. This means that 
the evolution proceeds at constant mass and constant luminosity. This has a very 
simple, but very important consequence for the relationship of gravity and 
effective temperature along each evolutionary track. From
\begin{equation}
L \propto R^{2}T^{4}_\mathrm{eff} = \mathrm{const.} ; M = \mathrm{const.}
\end{equation}
follows immediately that
\begin{equation}
M \propto g\;R^{2} \propto L\;(g/T^{4}_\mathrm{eff}) = L~g_{F} = \mathrm{const.}
\end{equation}

\begin{figure}[t]
\vskip-0.5 cm
\hskip-4mm 
 \includegraphics[width=0.75\linewidth,angle=90]{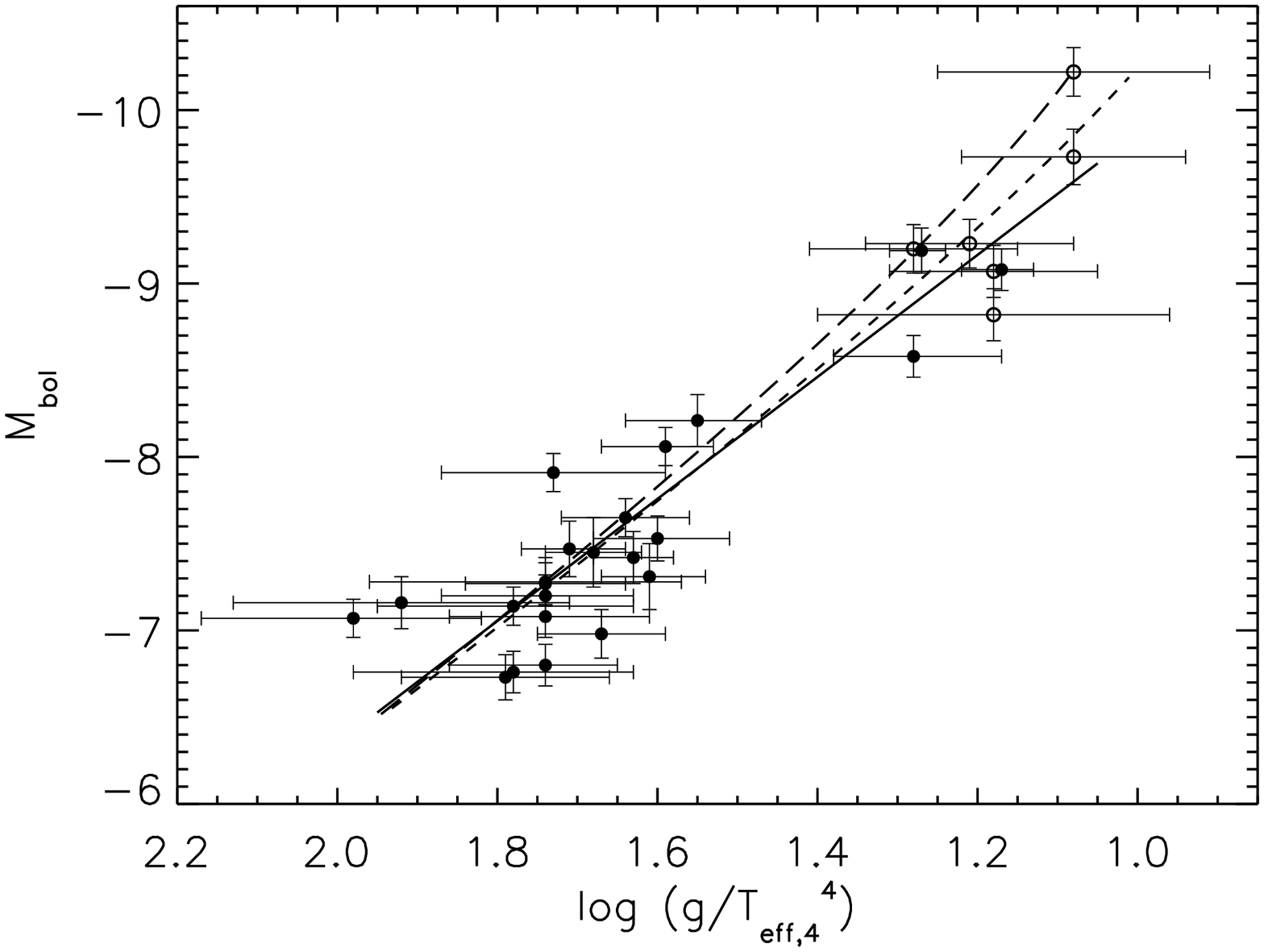} 
\caption{The FGLR of A (solid circles) and B (open circles) supergiants in 
NGC 300 and the linear regression (solid). The stellar evolution 
FGLRs for models with rotation are also overplotted 
(dashed: Milky Way metallicity, long-dashed: SMC metallicity).
}
   \label{fig16}
\end{figure}

Thus, along the evolution through the B and A supergiant domain the 
\emph{``flux-weighted gravity''} $g_{F} = g$/\teffq\/ should remain constant. 
This means each evolutionary track of different luminosity in this domain is characterized by a specific value of $g_{F}$. This value is determined by the 
relationship between stellar mass and luminosity, which in a first 
approximation is a power law
\begin{equation}
L \propto M^{x}\;
\end{equation}

and leads to a relationship between luminosity and flux-weighted gravity

\begin{equation}
L^{1-x} \propto (g/T^{4}_\mathrm{eff})^{x}\;.
\end{equation}

With the definition of bolometric magnitude 
$M_\mathrm{bol}$\,$\propto$\,$-2.5\log L$ one then derives

\begin{equation}
-M_\mathrm{bol} = a_{FGLR}(\log~g_{F} - 1.5)) + b_{FGLR}\;.
\end{equation}

This is the  \emph{``flux-weighted gravity -- luminosity relationship''} 
(FGLR) of blue supergiants. Note that the proportionality constant
$a_{FGLR}$ is given by the exponent of the mass -- luminosity power law through

\begin{equation}
a_{FGLR} = 2.5 x/(1-x)\;,
\end{equation}

for instance, for $x=3$, one obtains $a_{FGLR}=-3.75$. Note that the zero 
point of the relationship is chosen at a flux weighted gravity of 1.5, which is in the 
middle of the range encountered for blue supergiant stars.

\begin{figure*}[t]
 \includegraphics[width=0.73\textwidth, angle=90]{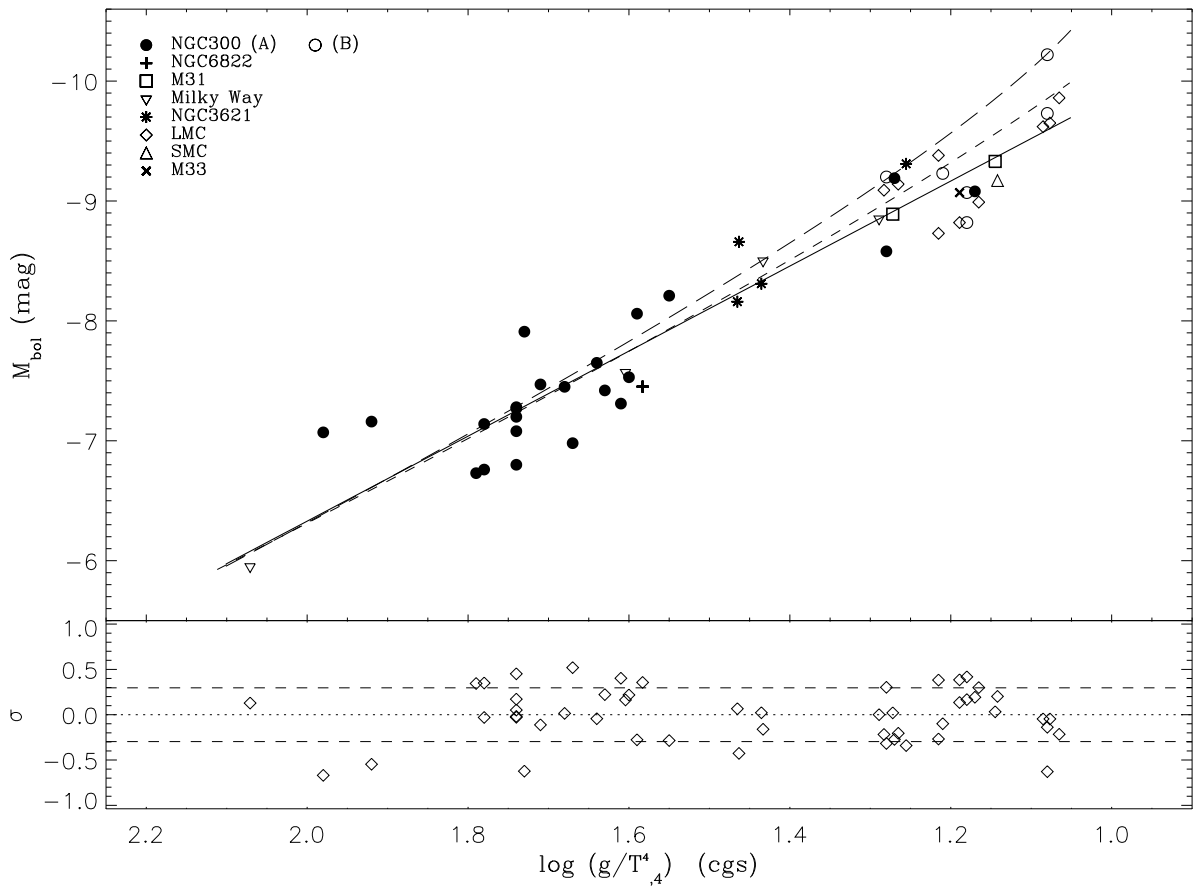} 
\caption{The FGLR of A (solid circles) and B (open circles) supergiants in 
8 galaxies including NGC 300 and the linear regression (solid). The stellar evolution 
FGLRs for models with rotation are again overplotted. 
}
   \label{fig17}
\end{figure*}

KUBGP use the mass-luminosity relationships of different evolutionary tracks
(with and without rotation, for Milky Way and SMC metallicity) to calculate the 
FGLRs predicted by stellar evolution. Very interestingly, 
while different evolutionary model types yield somewhat different FGLRs, 
the differences are rather small. 

\cite{kud03} were the first to realize that the FGLR has a very interesting 
potential as a purely spectroscopic distance indicator, as it relates two 
spectroscopically well defined quantitities, effective temperature and gravity, 
to the absolute magnitude. Compiling a large data set of spectroscopic high 
resolution studies of A supergiants in the Local Group and with an approximate 
analysis of low resolution data of a few targets in galaxies beyond the Local 
Group (see discussion in previous chapters) they were able to prove the 
existence of an observational FGLR rather similar to the theoretically 
predicted one.

With the improved analysis technique of low resolution spectra of A supergiants 
and with the much larger sample studied for NGC 300 KUBGP resumed the 
investigation of the FGLR.
The result is shown in Fig.\,\ref{fig16}, which for NGC 300 reveals 
a clear and rather 
tight relationship of flux weighted gravity $log~g_{F}$ with bolometric magnitude 
$M_{bol}$. A simple linear regression yields $b_{FGLR}$ = 8.11 for the zero point 
and 
$a_{FGLR}$ = -3.52 for the slope. The standard deviation from this relationship is 
$\sigma$ = 0.34 mag. Within the uncertainties the observed FGLR appears to be in 
agreement with the theory.

\begin{figure}[t]
\begin{center}
 \includegraphics[width=\linewidth]{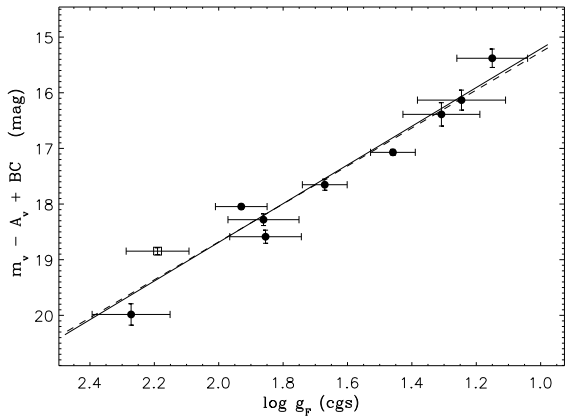} 
\caption{\small The FGLR of the Local Group dwarf irregular galaxy WLM, 
based on apparent bolometric magnitudes ($m_{bol}= m_v- A_v + BC$). The 
solid line corresponds to the FGLR calibration. The distance is determined 
by using this calibration through a minimization of the residuals. From 
\cite{urbaneja08}. 
}
   \label{fig18}
\end{center}
\end{figure}

In their first investigation of the empirical FGLR \cite{kud03} have added A supergiants 
from six Local Group galaxies with stellar parameters obtained from quantitative studies 
of high resolution spectra (Milky Way, LMC, SMC, M31, M33, NGC 6822) to their results 
for NGC 300 to 
obtain a larger sample. They also added 4 objects from the spiral galaxy NGC 3621 
(at 6.7 Mpc) which were studied at low resolution. KUBGP added exactly the same 
data set to their new enlarged NGC 300 sample, however, with a few minor modifications. 
For the Milky Way they included the latest results from \cite{przybilla06} and 
\cite{schiller08} and 
for the two objects in M31 we use the new stellar parameters obtained by 
\cite{przybilla06b}. For the objects in NGC 3621 they applied new HST photometry. 
They also re-analyzed the LMC objects using ionization equilibria for the 
temperature determination.

Fig.\,\ref{fig17} shows bolometric magnitudes and flux-weighted 
gravities for the
full sample of eight galaxies again revealing a tight relationship over one order of 
magnitude in flux-weighted gravity. The linear regression coefficients are 
$a_{FGLR} = -3.41\pm{0.16}$ and $b_{FGLR} = 8.02\pm{0.04}$, very simimilar to 
the NGC 300 sample 
alone. The standard deviation is $\sigma$ = 0.32 mag. The stellar evolution FGLR 
for Milky Way metallicity 
provides a fit of almost similar quality with a standard deviation of 
$\sigma$ = 0.31 mag.

\section{First Distances Using the FGLR-Method}

\begin{figure}[t]
\begin{center}
 \includegraphics[width=\linewidth]{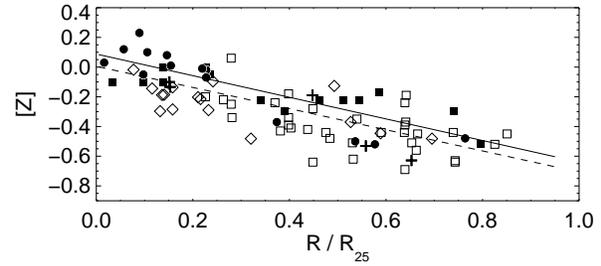} 

  \caption{Metallicity of blue supergiants, \hii\ regions and
      Cepheids as a function of dimensionless angular galactocentric
      distance in the triangulum galaxy M33. A supergiants (circles) 
      and B supergiants (squares) are shown as solid symbols.
      Logarithmic oxygen abundances of \hii\
      regions in units of the solar value as published by
      \cite{magrini07} are plotted as open squares. Logarithmic neon
      abundances of \hii\ regions normalized to the value for B
      stars in the solar neighbourhood and as obtained from
      \cite{rubin08} are shown as large open diamonds. The metallicity
      [$Z$] for beat Cepheids as determined by \cite{beaulieu06} are
      given as crosses. The solid line is the regression for the
      supergiants only, whereas the dashed lines is the regression for
      all objects. 
}

   \label{fig19}
\end{center}
\end{figure}

With a relatively small residual scatter of $\sigma \sim $ 0.3 mag the observed FGLR
with the calibrated values of $a_{FGLR}$ and $b_{FGLR}$  is an excellent 
tool to determine accurate spectroscopic distance to galaxies.
It requires multicolor photometry and low resolution ($5\AA$) spectroscopy to
determine effective temperature and gravity and, thus, flux-weighed gravity 
directly from the spectrum. With effective temperature, gravity and metallicity 
determined one also knows the bolometric correction, which is particularily 
small for A supergiants. This means that errors 
in the stellar parameters do not largely affect the determination of bolometric 
magnitudes. Moreover, one knows the intrinsic stellar SED and, therefore, can 
determine interstellar reddening and extinction from the multicolor photometry, 
which then allows for the accurate determination of the reddening-free apparent 
bolometric magnitude. The application of the FGLR then yields absolute 
magnitudes and, thus, the distance modulus.

\begin{figure}[t]
\begin{center}
 \includegraphics[width=\linewidth]{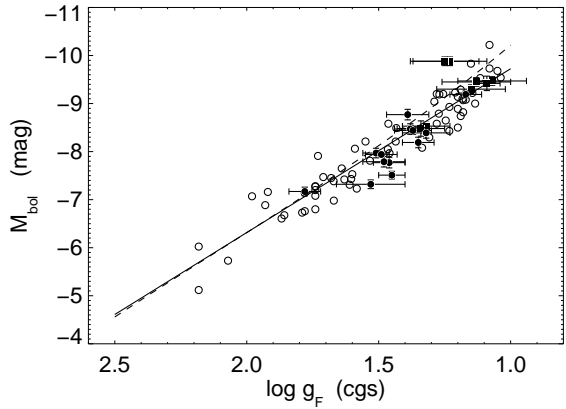} 
  
    \caption{FGLR fits of the blue supergiants in M33. Solid
      circles are late B and A supergiants and solid squares are early
      B supergiants in M\,33. The solid line is a linear fit as
      described in the text. In addition to the M\,33 targets,
      objects from nine other galaxies investigated in the studies by
      K08 and \cite{urbaneja08} are also shown. The solid line is the
      regression FGLR from K08. The dashed curve is the stellar
      evolution FGLR for Milky Way metallicity. 
   }

   \label{fig20}
\end{center}
\end{figure}

The first distance determination of this type has been carried out by 
\cite{urbaneja08} who studied blue supergiants in WLM, one of faintest dwarf irregular 
galaxies in the Local Group. The quantitative spectral analysis of VLT FORS spectra 
yields an extremely low metallicity of the young stellar population in this galaxy 
with an average of -0.9 dex below the solar value. The interstellar extinction is 
again extremely patchy ranging from 0.03 to 0.30 mag in E(B-V) (note that the 
foreground value given by \cite{schlegel98} is 0.037 mag). The individually de-reddened 
FGLR - in apparent bolometric magnitude - is shown in Fig.~\ref{fig18}. Using the FGLR 
calibration by \cite{kud08} and minimizing the residuals \cite{urbaneja08} determined a 
distance modulus of 24.99 $\pm0.10$ mag (995 $\pm46$ kpc). This value is in good agreement 
with the TRGB distance by \cite{rizzi07} and the K-band Cepheid distance by \cite{gieren08}, 
albeit 0.07 mag larger. 

Most recently, \cite{u09} have analyzed blue supergiant spectra obtained with 
DEIMOS and ESI at the Keck telescopes to determine a distance to the triangulum 
galaxy M33. The case of M33 is particulary interesting, since many independent 
distance determinations have been carried out for this galaxy during the last decade 
using a variety of techniques, including Cepheids, RR Lyrae, TRGB, red clump stars, 
planetary nebulae, horizontal branch stars and long-period variables. The surprising 
result of all of these studies has been that the distance moduli obtained with 
these different methods differ by as much as 0.6 mag, which is more than 30\% in 
the linear distance. At the same time, also the metallicity and the metallicity 
gradient of the young stellar population in the disk of M33 are still heavily disputed. 
Different published results obtained from \hii\ region work range from no gradient 
at all (\cite{rosolowsky08}) to a very steep gradient of -0.11 dex/kpc 
(\cite{garnett97}) and everywhere in between (\cite{willner02}, \cite{magrini07}) 
including a bimodal break with a very steep inner gradient(\cite{vilchez88}, 
\cite{magrini07}).

The blue supergiant spectroscopy by \cite{u09} supports a moderate metallicity 
gradient of 0.07dex/kpc without any indication of a bimodal break (see Fig.~\ref{fig19}).
The FGLR-method (Fig.~\ref{fig20})  yields a long distance modulus for M33 of 
24.93$\,\pm\,$0.11~mag, in basic agreement with a TRGB distance of
24.84$\,\pm\,$0.10~mag obtained by the same authors from HST ACS imaging. The long 
distance modulus agrees also very well with the eclipsing binary distance obtained 
by \cite{bonanos06}. \cite{u09} relate the difference between their result and the 
published cepheid distances to the difference in accounting for interstellar reddening 
(see Fig.~\ref{fig7}).

These successfull first applications of the FGLR-method for distance determinations 
indicate the very promising potential of the method to provide an independent constraint 
on the extragalactic distance scale.

\section{The Potential of FGLR-Method for extragalactic Distance Determinations}

One of the most challenging chapters of modern astrophysics is
the effort to establish the extragalactic distance scale with with sufficient 
accuracy. Over the past years, substantial improvement has
been made by the {\em HST} Key Project on the Extragalactic Distance Scale
(\cite{freedman01}) and by  the {\em SN Ia HST Calibration Program}
(\cite{saha01}, \cite{sandage06})  , in which Cepheid-based distances to
galaxies permitted the calibration of far-reaching secondary
indicators. In addition, the `Tip of the Red Giant Branch' (TRGB) 
method has become an additional very reliable and effective extragalactic 
distance indicator (\cite{rizzi07}), which can also be used to calibrate 
secondary indicators (Mould \& Sakai, 2008, 2009ab). However, in spite of 
this progress, there are still 
a number of systematic uncertainties which affect both, the Cepheids 
and secondary methods of distance measurement, and which do not yet 
allow to obtain extragalactic distances, and thus the Hubble constant, 
with the high accuracy desired and needed by cosmologists, i.e. below 
the current 10\% uncertainty. For instance, the results on the Hubble 
constant H$_{0}$ obtained by \cite{freedman01} and \cite{sandage06} differ 
by 20\% and the agreement with the TRGB-based calibration and the HST Key 
Project is at the margin of 10\%.  As is well known (see the discussion in 
Macri et al.~2006, section 4.6), the determination of cosmological 
parameters from the cosmic microwave background is affected by degeneracies 
in parameter space and cannot provide strong constraints on the value of 
H$_{0}$ (\cite{spergel06}, \cite{tegmark04}). Only, if assumptions are made, 
for instance that the universe is flat, H$_{0}$ can be predicted with high 
precision (i.e. 2\%) from the observations of the cosmic microwave background, 
baryionic accoustic oscillations and type I high redshift supernovae. If these 
assumptions are relaxed, then much larger uncertainties are introduced 
(\cite{spergel07}, \cite{komatsu09}). As an example, the uncertainty of the 
determination of the dark energy equation-of-state parameter w = p/(c$\rho^{2}$) 
is related to the uncertainty of the Hubble constant through 
$\delta$w/w $\sim 2 \delta$H$_{0}$/H$_{0}$. Thus, an independent 
determination of H$_{0}$ with an accuracy of 5\% will allow to reduce the 
1$\sigma$ uncertainty of the cosmological equation of state parameter $w$ 
to $\pm$0.1. A combination with other independent measurements constraining 
the cosmological parameters (large scale structure, SN Ia) will then allow 
for even tighter constraints. A very promising step forward in this regard has 
been made most recently by Riess et al., (2009ab), but it is clear that the 
complexity of their approach requires additional and independent tests.

Among the remaining uncertainties affecting the extragalactic distance 
scale, probably the most important one is interstellar
reddening. As young stars, Cepheids tend to be embedded in dusty
regions which produce a significant {\em internal} extinction, in
addition to the galactic foreground extinction. It seems likely that
most of the Cepheid distances to galaxies determined from optical
photometry {\em alone} are affected by sizeable systematic errors, due
to a flawed determination of the appropriate reddening
correction. The examples for NGC 300 and M33 give in the sections above 
clearly illustrate the problem. IR photometry (J, H and K-band) of Cepheids 
is a promising way to address the issue, as has been shown in the Araucaria 
collaboration (\cite{gieren05b}) or by \cite{riess09b}. However, photometry at 
these wavelengths is still subject to important systematic effects, thus an 
entirely independent and complementary approach to address the issue of reddening 
is highly desirable.

The TRGB method is, in principle, not 
free from reddening errors, too. Usually, fields in the outskirts of the 
galaxies investigated are observed and Galactic foreground reddening 
values obtained from the interpolation of published maps are used to 
apply an extinction correction. While this seems to be a reasonable 
assumption, it has also been demonstrated that for a few galaxies (NGC 300, 
\cite{vlajic09}; M83, \cite{bresolin09b}) the stellar disks 
extend much further out than previously assumed and what the intrinsic 
reddening is in these very faint extended disks is completely 
unexplored. In the application of the TRGB method, colors are used to constrain
metallicity and cannot, therefore, provide information about reddening.

An equally important uncertainty in the use of Cepheids as distance indicators is the
dependence of the period-luminosity (P-L) relationship on metallicity. Work 
by \cite{kennicutt98} on M101 and by \cite{macri06} on the 
maser galaxy NGC~4258 using metallicity gradient information from \hii\ region 
oxygen emission lines (Cepheids beyond the Local Group are too faint for a 
determination of their metallicity directly 
from spectra) indicates an {\em increase of Cepheid brightness with 
metallicity}. This agrees with \cite{sakai04}, who related the difference 
between TRGB and Cepheid distances to the galactic \hii\ region metallicities 
(not taking into account metallicity gradients, though) and derived a similar 
P-L dependence on metallicity. 

However, these results are highly uncertain. \cite{rizzi07} have argued that 
many of the TRGB distances used by \cite{sakai04} need to be revised. With the 
\cite{rizzi07} TRGB distances the \cite{sakai04} dependence of the P-L 
relationship on metallicity disappears and the results are in much closer agreement with 
stellar pulsation theory (\cite{fiorentino02}, \cite{marconi05}, \cite{bono08}), 
which predicts a small {\em decrease of Cepheid brightness with 
metallicity}. Moreover, all the \hii\ region (oxygen) metallicities adopted 
when comparing Cepheid distances with TRGB distances or when using metallicity 
gradients are highly uncertain. They result from the application of the 
``strong-line method'', using the calibration by \cite{zaritsky94}. As we have shown 
above, this calibration gives 
metallicities and metallicity gradients that are not in agreement with results 
obtained from blue supergiants or from \hii\ regions, when the more accurate method 
involving auroral lines is used. We point out, following the discussion in \cite{macri06}, 
that even small changes in the P-L metallicity dependence
can have an effect of several percent on the determination of H$_{0}$. We also note that the ,
most recent work by \cite{riess09b} makes use of this calibration.

Also the TRGB method has a metallicity dependence.
Usually, the metallicity of the old metal-poor population used for the method is 
obtained from the $V-I$ color, assuming that only foreground reddening is important. 
Then, a calibration of the TRGB magnitude as a function of metallicity is used. As  
shown in the careful work by \cite{mager08} on the TRGB distance to the maser 
galaxy NGC~4258, this metallicity correction introduces a systematic uncertainty of 
0.12 mag in the distance modulus.

Very obviously, in order to improve the determination of extragalactic 
distances of star-forming spiral and irregular galaxies in the local universe an 
independent and complementary method is desirable, which can overcome 
the problems of interstellar extinction and variations of chemical 
composition. The FGLR-method presented in the two previous sections is such a method. 
The tremendous advantage of this technique is that individual reddening and extinction 
values, together with metallicity, can be determined for each supergiant target 
directly from spectroscopy combined with photometry. This reduces significantly the 
uncertainties affecting competing methods such as cepheids and the TRGB.

The FGLR technique is robust. Bresolin et al.~2004, 2006 have shown, from observations 
in NGC\,300 and WLM, that the photometric variability of blue supergiants 
has negligible effect on the distances determined through the FGLR. 
Moreover, the study by \cite{bresolin05} also confirms 
that with {\em HST} photometry the FGLR method is not affected by crowding out to 
distances of at least 30 Mpc. This is the consequence of the enormous intrinsic 
brightness of these objects, which are 3 to 6 magnitudes brighter than Cepheids.

The current calibration of the FGLR rests on the 5\,\AA\ resolution stellar spectra 
obtained in NGC~300 and in seven additional galaxies (in some cases with higher resolution). 
Ideally, a large number of stars, observed in a single galaxy 
with a well-established distance, should be used. Obviously, a very natural step to 
improve the precision of the method by recalibrating in the LMC. The LMC currently 
defines the zero point for many classic photometric distance indicators, and its
distance is presently well constrained, e.g.~from eclipsing binaries 
($m-M=18.50\pm0.06$, \cite{pietrzynski09}).
This work is presently under way. In addition, the analysis of a large sample of 
SMC blue supergiants will provide additional information about the metallicity dependence 
of the method, but also about the geometrical depth of the SMC. New spectra in IC~1613 
(10\% solar metallicity) and M31 ($\sim$ solar metallicity) will help to constrain how 
the FLRG depends on metallicity, even though the results already obtained on WLM 
(10\% solar metallicity, \cite{urbaneja08}) does not indicate a strong effect.

An alternative and certainly more ambitious approach will be to by-pass the LMC as a distance 
scale anchor point and to use the maser galaxy NGC 4258 (\cite{humphreys08}) as the 
ultimate calibrator. While challenging, such a step would be entirely feasible and 
provide an independent test of the approach taken by Riess et al. (2009ab). Last but not 
least, the GAIA mission will provide a very accurate determination of the distances of blue 
supergiants in the Milky Way, which will allow for an accurate local calibration of the FGLR.

\section{Perspectives of Future Work}

It is evident that the type of work described in this paper can be in
a straightforward way extended to the many spiral galaxies in the
local volume at distances in the 4 to 12 Mpc range. \cite{bresolin01}
have already studied A supergiants in NGC 3621 at a distance of 7 Mpc.
Pushing the method we estimate that with present day 8m to 10m class
telescopes and the existing very efficient multi-object spectrographs
one can reach down with sufficient S/N to V = 22.5 mag in two nights
of observing time under very good conditions. For objects brighter
than $M_{V}$ = -8 mag this means metallicities and distances can be
determined out to distances of 12 Mpc (m-M = 30.5 mag). This opens up
a substantial volume of the local universe for metallicity and
galactic evolution studies and independent distance determinations
complementary to the existing methods. With the next generation of
extremely large telescopes such as the TMT, GMT or the E-ELT the
limiting magnitude can be pushed to V = 24.5 equivalent to distances
of 30 Mpc (m-M = 32.5 mag).

\acknowledgements
The work presented here is the result of an ongoing collaborative
effort over many years.  I wish to thank my colleagues in Hawaii
Miguel Urbaneja, Fabio Bresolin and Vivian U for their tremendous
dedication and skillful contributions to help me to make this project
happen. It is more than just a science collaboration. It is a joyful
endeavor with a lot of challenges but also a lot of fun. My colleagues
Norbert Przybilla and Florian Schiller from Bamberg Observatory have
made crucial contributions without which this project would not have
been possible. I hope that the good time we had together in Hawaii has
been an adequate compensation for their efforts. I also want to thank
Wolfgang Gieren and his team from Universidad de Concepcion for
inviting us to be part of the Araucaria collaboration. This has given
a much wider perspective to our project. It has also created the new
spirit of a ``trans-Pacific'' collaboration.


\end{document}